# Understanding the Landscape of Leveraging IoT for Sustainable Growth in Saudi Arabia

Manal Alshehri[1], and Ohoud Alharbi[1]

[1]King Saud University, Riyadh, KSA

445920919@student.ksu.edu.sa, and omalharbi@ksu.edu.sa

**Abstract**

The integration of Internet of Things (IoT) technologies in agriculture holds promise for transforming farming practices, particularly in the Kingdom of Saudi Arabia (KSA). This study explores the adoption of smart farming practices among KSA farmers. Due to the geographical location and nature of KSA, it faces significant challenges in agriculture. The objective of this research is to discuss how IoT will enhance agriculture in KSA and identify its current usage by conducting a study on Saudi farmers with varying ages, regions, and years of experience. The results indicate that 90% of the farmers encounter challenges in farming, and all of them express interest in adopting smart farming to address these issues. While 60% of farmers are currently utilizing IoT technologies, they encounter challenges in implementing smart farming practices. Thus, smart farming presents solutions to prevalent challenges including adverse weather, water scarcity, and labor shortages, though barriers include cost and educational challenges.

1. **Introduction**

The Internet of Things (IoT) is the concept of physical objects being connected through the Internet to exchange data with other devices and systems without human intervention, establishing a computer-to-computer relationship [11, 14]. This concept has recently gained significant popularity, with a sharp increase in the number of applications utilizing this technology across various domains such as smart homes, healthcare, and knowledge management [12]. Agriculture has emerged as a recent adopter of IoT technology, showcasing promising results. Enhancing agricultural practices in Saudi Arabia is a key target of the country's 2030 vision, aiming to improve the quality of life by implementing sustainable urban greening initiatives. However, agriculture in Saudi Arabia faces significant challenges due to its arid nature, resulting in limited water resources [4], and its vulnerability to climate change factors due to its geographical location [2].

This research endeavors to investigate the applications of IoT technology in agriculture through the utilization of various tools and techniques. Subsequently, it discusses the potential of IoT in enhancing agriculture in Saudi Arabia and proposes solutions for agricultural applications in the region using IoT technology. This study seeks to address the following research questions:

- *How are IoT technologies being used in agriculture in Saudi Arabia?*
- *What are the challenges encountered when implementing IoT technologies in the agricultural sector in Saudi Arabia?*

## 2. Related Work

While software engineering research in Saudi Arabia predominantly concentrates on tools to enhance workplace productivity [24], AI integration into development tools [25], the introduction of new technologies or interaction styles [23], and the study of navigation and robotics [26,27], there remains a notable scarcity of research focusing on the utilization of IoT in agriculture in Saudi Arabia.

The Internet of Things (IoT) can be defined as devices connected through the Internet to provide services [14]. It has been instrumental in supporting agricultural applications, aiding in farm monitoring and management, including water, irrigation, soil, and weather management [6, 10, 15]. Numerous studies explore IoT applications in agriculture, detailing tools used and evaluating each application. Challenges faced by IoT in agriculture are also discussed.

One study proposed a prototype enabling farmers to monitor their farms via phones, presenting real-time data. The system includes a data store for real-time data collected from sensors measuring greenhouse temperature, light, pressure, and humidity. Communication utilizes MicaZ wireless modules programmed in nesC, with TinyOS as the operating system [1]. While data is immediately available in real-time and stored briefly, it can also be stored long-term for analysis.

Another application introduces an intelligent system for monitoring greenhouse temperature and light, regulating temperature and controlling light to maintain optimal conditions for plant growth. The system also handles data analysis for agriculture-related predictions [16].

Another study proposed models for implementing smart agriculture through cloud management and security. One model is a simple IoT system monitoring farm temperature and humidity, providing farmers with future prediction details. Deep learning algorithms were employed to develop an IoT framework for accurate weather predictions based on collected data [9].

Systems have been developed for controlling and managing plant irrigation. These systems use sensors to measure moisture and monitor crops, sending data through Wi-Fi modules to a database via a web server. Decisions regarding irrigation are made based on sensor data, with dashboards provided for monitoring and control [8].

Other systems focus on real-time decision-making based on stored data, incorporating weather data and sensor-collected data via wireless connectors. These web-based systems enable farmers to manage their farms, share knowledge, and collaborate. However, they may face limitations in large-scale implementation [5].

Challenges in IoT agriculture include implementation and maintenance costs, sensor manufacturing compatibility with technologies and networking, varied hardware maintenance solutions, environmental impacts of devices, and security and privacy concerns associated with sharing information over the Internet [3, 6, 7, 9, 13, 15, 17, 19–22]. Additionally, real-time monitoring systems and data management add to the cost [7, 13, 15, 17, 19–22].

After reviewing research papers about the applications of IoT in agriculture and the challenges faced when using IoT, we concluded that IoT is mostly used in three main categories: controlling, monitoring, and managing farms.

## 3. Methodology

To investigate the prevalence and awareness of smart farming among Saudi Arabian farmers, a survey was conducted with the facilitation of the research unit of the Ministry of Agriculture in Saudi Arabia. The survey aimed to gauge the knowledge and awareness of Saudi farmers regarding the adoption of modern technologies in agriculture. It consisted of two sections: general questions and detailed questions. The general questions section collected demographic information, including gender, age, years of experience in farming, region of residence, level of education, and agricultural field specialization. The detailed questions section focused on assessing farmers' perceptions and practices related to smart farming and modern agricultural technologies. Questions covered topics such as challenges faced in agriculture, awareness of smart farming and IoT, usage of sensors for field monitoring, opinions on the effectiveness and popularity of smart farming, and intentions to adopt new technologies in farming practices.

Farmers were provided with the survey electronically. Participation in the survey was voluntary and anonymous, and respondents were assured that their responses would be used solely for research purposes.

Upon completion of data collection, responses were compiled and analyzed to discern patterns and trends regarding the popularity and awareness of smart farming among Saudi Arabian farmers. Descriptive statistics were employed to summarize demographic information and survey responses, providing insights into the current landscape of agricultural technology adoption in the Kingdom of Saudi Arabia.

## 4. Result

The chosen participants were 10 Saudi male farmers with a variety of ages, regions, and years of experience, as shown in Figures 1,2,3 and 4. They were asked before filling out the survey to agree to use their filled answers for research purposes. The majority of participants are older than 45 years old and with more than 10 years of experience (see Figures 1 and 2). There is diversity in the regions, see Figure 3. In addition, the background of education varies, but all of the participants have at least a diploma.

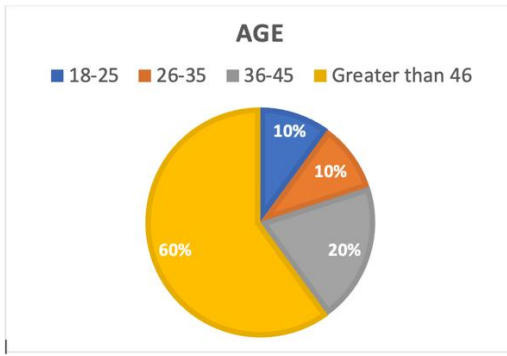
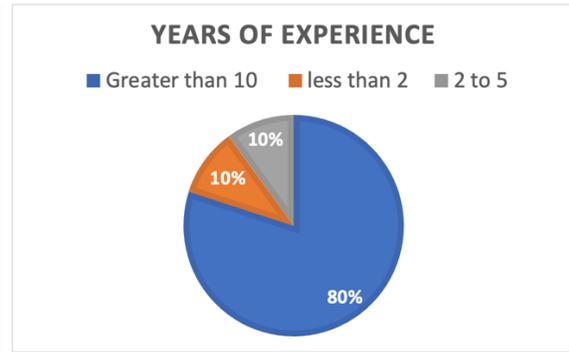

*Figure 1 Ages of participants*  *Figure 2 Years of experience for participants*

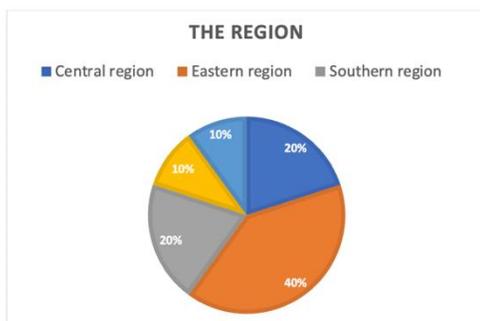
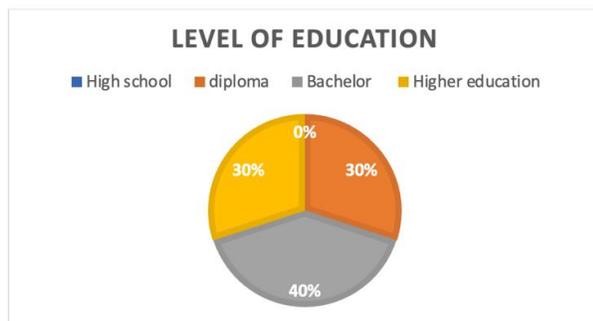

*Figure 3 The region of participants*  *Figure 4 Educational background of participants*

Most farmers face challenges in agriculture, and climate change is the most challenging factor in KSA. As a result, most farmers track the weather daily (see Figure 5).

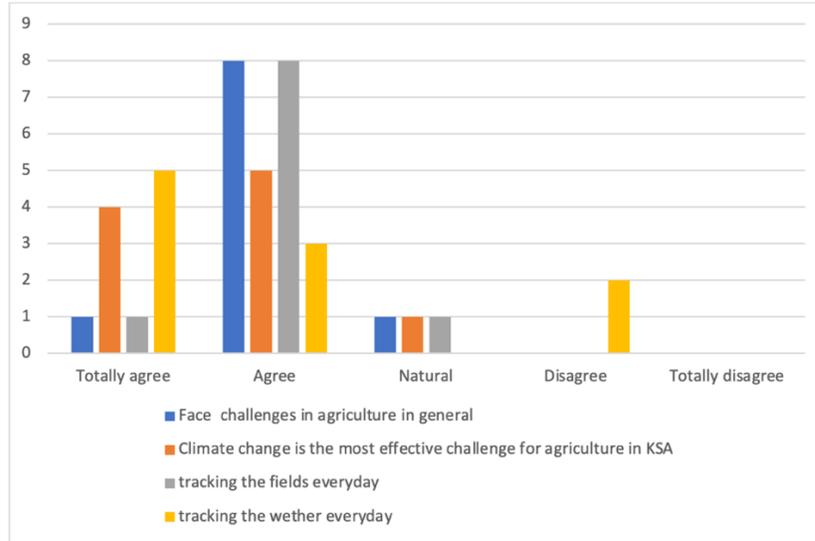

*Figure 5 Challenges and tracking*

The participants were asked about their knowledge and awareness of smart farming (see Figure 6). The findings indicate that 60% of participants are currently applying the latest technologies in farming, such as using sensors and automatic irrigation. Growing fruits and vegetables is the most common field applying IoT and the latest technologies in agriculture.

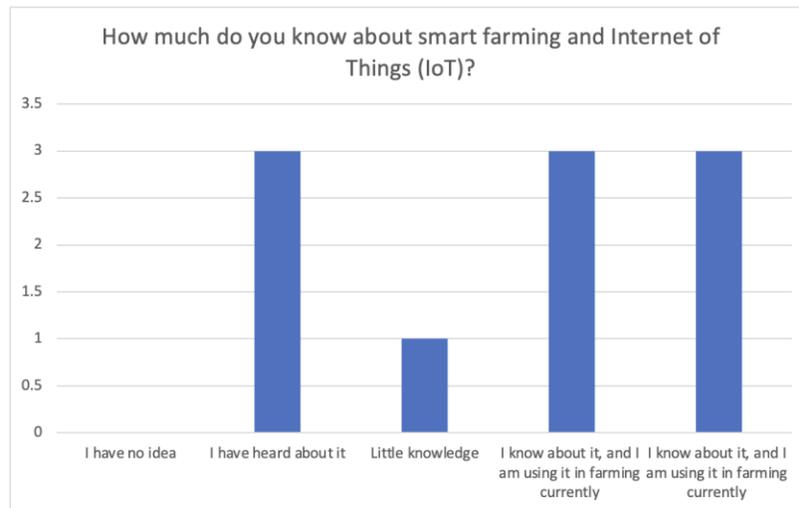

*Figure 6 Farmers' knowledge of smart farming*

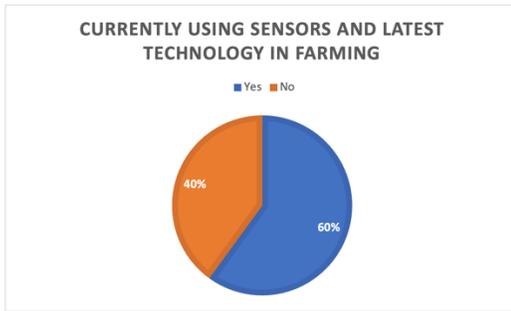 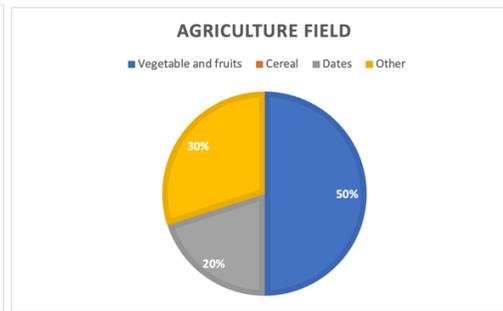

*Figure 7 Currently using IoT in their farms*   *Figure 8 The fields used the IoT*

Results show there is popularity for using smart farming in KSA, and most farmers agree that it helps in increasing crop production (see Figure 9). The farmers were asked if they would apply these technologies for smart farming in their farms if given the chance, and all answers were "Yes" with 100%.

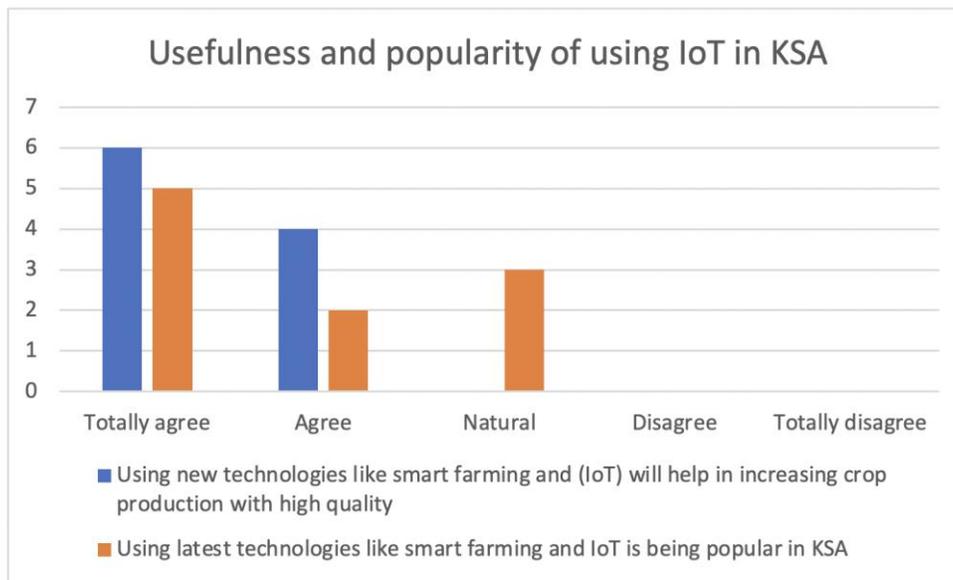

*Figure 9 Usefulness and popularity of using IoT in KSA*

Even though it is popular and helpful for farmers, there are some challenges in applying these new technologies. These include the high cost of some sensors and machines, difficulty in learning new technologies, and maintenance challenges (see Figure 10).

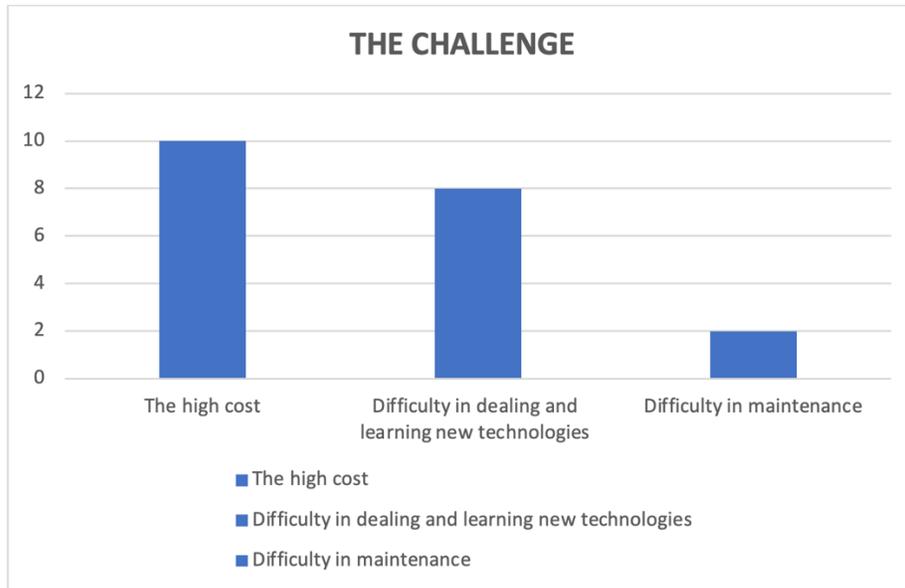

*Figure 10 Challenges faced when applying smart farming*

To summarize, 90% of the farmers face challenges in farming, and all of them would like to apply smart farming if given the chance to solve these problems. 60% of farmers are currently using these technologies, but they are facing other challenges in applying smart farming. All of them agreed that the high cost of sensors and tools is the most significant challenge.

5. Discussion

The study surveyed Saudi farmers with diverse backgrounds in terms of age, region, and years of experience, as illustrated in Figures 1, 2, 3, and 4. The majority of participants were older than 45 years old and possessed more than a decade of farming experience, indicating a wealth of practical knowledge (Figures 1 and 2). Notably, the participant pool represented various regions of Saudi Arabia, as depicted in Figure 3. Moreover, while educational backgrounds varied, all participants held at least a diploma, ensuring a baseline level of education across the sample. We found that there is a trend between education and the application of smart farming.

The findings underscored the prevalent challenges encountered by farmers in Saudi Arabia, with climate change emerging as the most formidable obstacle. Consequently, a majority of farmers reported tracking weather patterns daily, as depicted in Figure 5. Participants demonstrated a substantial level of awareness of smart farming, with 60% currently employing advanced technologies such as sensors and automatic irrigation (Figure 6). IoT appears to be popular in KSA, particularly in the eastern region. Notably, the cultivation of fruits and vegetables emerged as the primary field leveraging IoT and modern agricultural technologies (Figure 8), despite dates being the primary agricultural product in KSA according to the General Authority of Statistics [18].

The problems faced by farmers are related to weather, lack of water sources, dry soil, and a shortage of labor. These problems can potentially be alleviated using smart farming technologies. The table below lists the problems and suggested solutions:

| Problems | Suggested solutions |
|---|---|
| Weather | Applying greenhouse solutions with temperature monitoring by sensors |
| Lack of water source | Limiting water waste by controlling the irrigation process; modern sensors in a simple system would work well. |

| Impact of insects | Protecting fields in greenhouses would prevent birds from eating the plants |
|---|---|
| Lack of labor and pressure on them | Technologies could eliminate the work needed by humans. |

Results further indicated the widespread popularity of smart farming among Saudi farmers, with a consensus that it contributes to increased crop production, as depicted in Figure 9. Encouragingly, all participants expressed willingness to adopt these technologies if given the opportunity, underscoring their perceived utility and acceptance (Figure 9). However, despite its benefits, challenges persist in the implementation of smart farming technologies, notably the high cost of sensors and equipment, as well as difficulties in learning and maintenance (Figure 10).

The most reported challenge, according to the collected data, is the high cost, which aligns with findings from the literature review. The second challenge is the difficulty in dealing with and learning new technologies, despite the majority of participants having at least diploma degrees. On the other hand, only a few farmers mentioned difficulty in maintenance, whereas maintenance is considered one of the most significant challenges in previous studies [3,7,9].

A major limitation of this study was the lack of reliable research available on agriculture in the Kingdom of Saudi Arabia (KSA). Moreover, the survey had a relatively small number of participants, with only 10 farmers involved. However, despite these limitations, the exploratory nature of the survey study allowed for valuable insights to be gained from the firsthand perspectives of the participating farmers. While the small sample size may not fully capture the diversity of experiences across all regions and agricultural sectors within KSA, it still provides preliminary data and insights that can inform future research efforts. Therefore, while the limitations are acknowledged, they do not significantly detract from the exploratory nature and initial insights provided by the survey study.

In conclusion, the study reveals a strong inclination towards adopting smart farming practices among Saudi farmers, driven by their recognition of its potential to address agricultural challenges and enhance productivity. While obstacles exist, such as cost and technical complexities, the overwhelmingly positive attitude towards smart farming signals promising opportunities for its integration into Saudi agricultural practices.

## 6. Conclusion

The integration of Internet of Things (IoT) technologies marks a significant evolution in traditional farming methods in Saudi Arabia, presenting unparalleled opportunities for optimization and advancement. This study presents a comprehensive overview of Saudi farmers' perspectives on smart farming practices, revealing a notable interest and awareness among participants despite facing significant challenges. The study provides valuable insights into the current landscape of smart farming adoption in the Kingdom. The findings underscore the potential of smart farming technologies to address pressing agricultural issues, including climate change, water scarcity, and labor shortages. However, barriers such as high costs and technological complexities remain obstacles to widespread adoption. Nonetheless, the overwhelmingly positive attitude towards smart farming signals promising opportunities for its integration into Saudi agricultural practices, highlighting the need for further research and targeted interventions to support its implementation and maximize its benefits for the agricultural sector.